\documentclass[12pt,a4paper]{article}
\usepackage{layout}
\usepackage{amsmath}
\usepackage{textcomp}
\usepackage{graphicx}
\usepackage[left=1.5cm,top=1.cm,right=1cm]{geometry}
\date{}

\begin{document}
\title{Coupled tensorial form for atomic relativistic two-particle operator given in second quantization representation}
\maketitle




\author{Rytis Jur\v{s}\.{e}nas,
        Gintaras Merkelis}

     {Institute of Theoretical Physics and Astronomy of Vilnius University,\\
     A. Go\v{s}tauto 12, LT-01108 Vilnius, Lithuania}



\abstract{General formulas of the two-electron operator representing either
atomic or effective interactions are given in a coupled tensorial
form in relativistic approximation. The alternatives of using uncoupled,
coupled and antisymmetric two-electron wave functions in constructing
coupled tensorial form of the operator are studied. The second quantization
technique is used. The considered operator acts in the space of states of open-subshell
atoms.}





\section{Introduction}

In the atomic structure calculations, investigations to optimize
the effort of obtaining matrix elements of a two-electron operator
are urgently required. This can be explained by the fact that theoretical
methods recently used to produce high-precision atomic structure data
generate large sets of matrix elements for a two-particle
operator. This leads to large computation requirements in terms of both
memory and speed. For example, large-scale configuration interaction
(CI) calculations \cite{cfflarg}-\cite{ggfrgrant} use a massive matrix for the atomic
Hamiltonian. A large fraction of expansion terms of perturbation theory
(PT) recently applied in atomic calculations \cite{Lin82}-\cite{mkrf85} are considered
as matrix elements of some effective two-particle operators with complex
tensorial structures. In all the above mentioned studies, a significant fraction
of computations are devoted to the calculation of $N$-electron angular
parts of the matrix elements of a two-electron operator. Particularly
complex calculations emerge when more than two open subshells with
$N>2$ are involved. A number of methods and techniques
\cite{ja73}-\cite{mg20} were developed in order to obtain the general formulas for matrix
elements of a two-particle operator for many electron case. The comprehensive
description of this subject can be found everywhere \cite{ja73},
\cite{zr97}.

In the present paper we distinguish the second quantization representation
(SQR) \cite{Ho}, \cite{J63}, \cite{RK84}. The efficiency of this
technique manifests itself when the tensorial properties of creation and
annihilation operators are taken into account \cite{RK84}. Then the
$N$-electron angular part of a matrix element is described by a coupled
tensorial product of creation and annihilation operators. In order
to optimize (minimize) the calculation procedures it is important
to choose the appropriate coupling schemes of angular momenta and
the order of creation and annihilation operators in the tensorial
product. In \cite{ggfrgrant} and \cite{grcff} the coupling schemes
for the tensorial products of creation and annihilation operators
were considered for nonrelativistic ($LS$-coupling) and relativistic
($jj$-coupling) cases, respectively. The manner to determinate the
expressions for matrix elements was presented. In \cite{mg20} a
coupled tensorial form of an effective two-particle operator used
in a second-order MBPT was obtained in $LS$-coupling. Here, the different
forms \cite{grcff} of coupling schemes to make tensorial product
for particular cases were suggested. This enables one to reduce the complexity
of the expressions for matrix elements. In \cite{jurm2007} the investigations
of \cite{mg20} were extended by including into the presentation of
coupled tensorial form of a two-particle operator coupled and antisymmetric
two-electron wave functions given in $LS$-coupling.

In the present manuscript we continue the studies of \cite{jurm2007} by
considering a two-particle operator in $jj$-coupling (in the relativistic
approximation \cite{parpia}). We search for the general expressions
of formal (effective) two-particle operator which describes atomic
interactions as well as effective interactions in atoms. In Section
\ref{sc:sc2} a coupled form of the two-electron operator is studied using
uncoupled, coupled and antisymmetric two-electron wave functions.
Sets of expressions for the two-electron operator are given in SQR (see Tables).
An example of the application of obtained results for specific cases is
considered in Section \ref{sc:sc3}.


\section{Coupling schemes of ranks for a two-particle operator}\label{sc:sc2}

Let us consider a two-particle operator $G$ given in the second quantization
representation (SQR) in $jj$-coupling \cite{RK84}:
\begin{equation}
G=X\displaystyle \sum_{\alpha \beta \mu \nu} a_{\alpha}a_{\beta}a_{\nu}^{\dagger}a_{\mu}^{\dagger} \, \, g(\alpha,\beta,\mu,\nu;\gamma,m_{\gamma}).\label{ggr}\end{equation}
In our considerations $\alpha,\beta,\nu,\mu$ indicate the subshells $n_{\alpha}\lambda_{\alpha}m_{\alpha},n_{\beta}\lambda_{\beta}m_{\beta},n_{\nu}\lambda_{\nu}m_{\nu},n_{\mu}\lambda_{\mu}
m_{\mu}$
of $N$-electron wave function ${{\left\vert \Psi^{N}\right\rangle \equiv\left\vert n_{a}\lambda_{a}^{N_{a}}\Lambda_{a}\, n_{b}\lambda_{b}^{N_{b}}\Lambda_{b}...n_{k}\lambda_{k}^{N_{k}}\Lambda_{k}\,\left(\Lambda_{ab}...\right)\;\Lambda M_{\Lambda}\right\rangle }}$,
the operator $G$ acts on. Operators $a_{i}$ and $a_{i}^{\dagger}$
denote electron creation and annihilation operators in the state $n_{i}\lambda_{i}m_{i}$
($\lambda_{i}=l_{i}j_{i}$, $l_{i}-$parity of the state) with principal quantum number $n_{i}$ and magnetic quantum number (projection) $m_{i}$. In the present paper the factor $g(\alpha,\beta,\mu,\nu;\gamma,m_{\gamma})$
is associated with a matrix element ${\left\langle n_{\alpha}\lambda_{\alpha}m_{\alpha}n_{\beta}\lambda_{\beta}m_{\beta}\left\vert g_{m_{\gamma}}^{(\gamma)}(1,2)\right\vert n_{\mu}\lambda_{\mu}m_{\mu}n_{\nu}\lambda_{\nu}m_{\nu}\right\rangle }$
of a two-electron operator\begin{equation}
G=\displaystyle \sum_{i<j}^{N} g_{ij}=\frac{1}{2}\sum_{i\neq j}\, g_{ij}\label{gbigkoord}\end{equation}
of atomic interactions. We assume that $g_{ij}$ can be expressed
by the tensorial product \cite{ja73}:
\begin{equation}
g_{ij}\equiv g_{m_{\gamma}}^{(\gamma)}(i,j)=\displaystyle \sum_{\gamma_{1}\gamma_{2}} g(r_{i},r_{j})g_{m_{\gamma}}^{(\gamma_{1}\,\gamma_{2})(\gamma)}=\displaystyle \sum_{\gamma_{1}\gamma_{2}} g(r_{i},r_{j})\left[g^{(\gamma_{1})}(i)\times g^{(\gamma_{2})}(j)\right]_{m_{\gamma}}^{(\gamma)}\label{gcr}
\end{equation}
and that $g_{ij}=g_{ji}.$ In (\ref{gcr}) $g(r_{i},r_{j})$ is the
radial part of $g_{ij}$. An irreducible tensorial operator $g^{(\gamma)}(i)$
acts on the spin-angular variables of the $i$-th
electron in the space of one-electron relativistic wave functions
($4$-spinors) \cite{zr97} \begin{equation}
\left\vert n\lambda m\right\rangle \equiv\left\vert nljm\right\rangle =\left(\begin{array}{c}
f(nlj|r)\left\vert ljm\right\rangle \\
(-1)^{\vartheta}g(nl^{\prime}j|r)\left\vert l^{\prime}jm\right\rangle \end{array}\right)\label{onerelfunk},\end{equation}
where $l^{\prime}=2j-l$, $\vartheta=l-j+1/2$. Functions $f(nlj|r)$
and $g(nl^{\prime}j|r)$ are large and small components of $\left\vert n\lambda m\right\rangle$.
Functions $\left\vert ljm\right\rangle$, $\left\vert l^{\prime}jm\right\rangle$ are $2$-spinors. In (\ref{ggr}) the factor
$X=1/2$. However, when $g(\alpha,\beta,\mu,\nu;\gamma,m_{\gamma})$
is associated with the antisymmetric matrix element \cite{johnson1}
\[
\left\langle n_{\alpha}\lambda_{\alpha}m_{\alpha}n_{\beta}\lambda_{\beta}m_{\beta}\left\vert g_{m_{\gamma}}^{(\gamma)}(1,2)\right\vert n_{\mu}\lambda_{\mu}m_{\mu}n_{\nu}\lambda_{\nu}m_{\nu}\right\rangle _{A}\]
\begin{equation}
=\left[1-\left(\mu \leftrightarrow \nu\right)\right]\left\langle n_{\alpha}\lambda_{\alpha}m_{\alpha}n_{\beta}\lambda_{\beta}m_{\beta}\left\vert g_{m_{\gamma}}^{(\gamma)}(1,2)\right\vert n_{\mu}\lambda_{\mu}m_{\mu}n_{\nu}\lambda_{\nu}m_{\nu}\right\rangle,\label{eq:antap}
\end{equation}

\noindent the factor $X=1/4$. Here $(\mu\leftrightarrow\nu)$ indicates that $\mu$ must be interchanged with $\nu$.

Notice, that when examining the atomic perturbation theory expansion terms
or coupled cluster (CC) approach equation ones, they can be considered
as the matrix elements of some effective operator $^{eff}g_{m_{\gamma}}^{(\gamma)}$ \cite{Lin82}.
Then the factor $g(\alpha,\beta,\mu,\nu;\gamma,m_{\gamma})$ can be
associated with the matrix element ${\left\langle \alpha\beta\left|^{eff}g_{m_{\gamma}}^{(\gamma)}\right|\mu\nu\right\rangle }$.
The operator $^{eff}g_{m_{\gamma}}^{(\gamma)}$ usually has more complicated
tensorial structure and symmetry properties than $g_{m_{\gamma}}^{(\gamma)}$
(\ref{gcr}). Nevertheless, the expressions of $G$ developed in this
manuscript are also valid for $^{eff}g_{m_{\gamma}}^{(\gamma)}$. In the
later case the factor $X$ is obtained individually.

\begin{figure}
\framebox[10cm]{\includegraphics[scale=0.7]{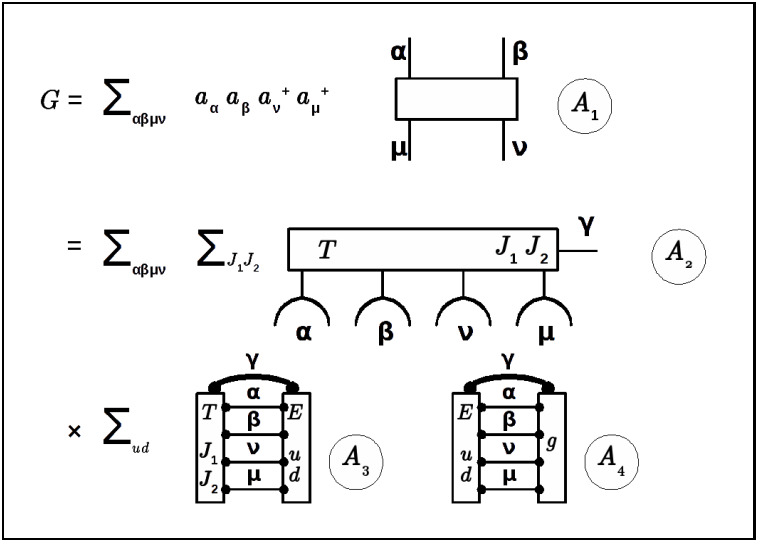}}
\caption{The block-scheme of transforming two-electron operator into coupled tensorial form.}
\label{fig:Figure1}
\end{figure}

Below we briefly describe the procedures that we have used to convert
operator $G$ into a coupled tensorial form, \textit{i.e.}, to obtain
the expressions for $G$, where the quantities entering these expressions
are independent of magnetic quantum numbers $m_{i}$. It is
convinient to explain such transformation by examining the schema
(Figure \ref{fig:Figure1}) which arises when applying a graphical method of angular
momentum theory \cite{JB65}. In schematic form we can write
\begin{equation}
G=X\displaystyle \sum_{\alpha\beta\mu\nu} \,\, \sum_{m_{\alpha}m_{\beta}m_{\mu}m_{\nu}}a_{\alpha}a_{\beta}a_{\nu}^{\dagger}a_{\mu}^{\dagger}\, A_{1}=\sum_{\alpha\beta\mu\nu}\, \sum_{J_{1}J_{2}}\ A_{2}\,\sum_{ud}\, A_{3}\,\ A_{4},\label{sscchema}
\end{equation}


where summation indices $\alpha\beta\mu\nu$ denote quantum numbers $n\lambda$. The diagram $A_{1}$ denotes either a matrix element ${\left\langle \alpha\beta\left|g_{m_{\gamma}}^{(\gamma)}\right|\mu\nu\right\rangle }$
or ${\left\langle \alpha\beta\left|^{eff}g_{m_{\gamma}}^{(\gamma)}\right|\mu\nu\right\rangle }$. The irreducible tensorial product (the diagram $A_{2}$) composed
of creation and annihilation operators was produced by using Jucys
theorems of graphical angular momentum theory \cite{JB65}. To obtain
the desired form of the irreducible product, we carried out several
recouplings of angular momenta and made several changes of positions
of creation and annihilation operators in (\ref{ggr}). The arrangements
of operators will be discussed later. Firstly, we shall discuss the recoupling
of angular momenta. In this manuscript we investigate two approaches. In
each approach the specific coupling schemes of the angular momenta
$j$ were applied. The block $E$ of diagrams $A_{3}$, $A_{4}$ represents
such schemes. In the first approach we have used the following coupling
of momenta\begin{equation}
E_{b}:=\left[\begin{array}{ccc}
j_{\mu} & u & j_{\alpha}\\
m_{\mu} & m_{u} & m_{\alpha}\end{array}\right]\left[\begin{array}{ccc}
j_{\nu} & d & j_{\beta}\\
m_{\nu} & m_{d} & m_{\beta}\end{array}\right]\left[\begin{array}{ccc}
u & d & \gamma\\
m_{u} & m_{d} & m_{\gamma}\end{array}\right],\label{ebscem}\end{equation}
while in the second approach, the block $E$ is given by \begin{equation}
E_{z}:=\left[\begin{array}{ccc}
j_{\alpha} & j_{\beta} & u\\
m_{\alpha} & m_{\beta} & m_{u}\end{array}\right]\left[\begin{array}{ccc}
j_{\mu} & j_{\nu} & d\\
m_{\mu} & m_{\nu} & m_{d}\end{array}\right]\left[\begin{array}{ccc}
u & d & \gamma\\
m_{u} & m_{d} & m_{\gamma}\end{array}\right].\label{ezschem}\end{equation}
Here $\left[...\right]$ brackets define Clebsch-Gordan coefficients; $u$ and $d$ are the intermediate momenta arising in the recoupling
procedure. The block $T$ in $A_{2}$, $A_{3}$ describes the coupling
schema of the irreducible tensorial product of creation and annihilation
operators with the intermediate ranks $J_{1}$, $J_{2}$. The diagram
$A_{3}$ represents the recoupling coefficient of angular momenta
$j_{i}$ transforming the schema $E$ into $T$. In the case of (\ref{gcr}),
due to the chosen specific coupling $E_{b}$, in the first approach
the diagram $A_{4}$ corresponds to product of the submatrix element
$\left[\lambda_{\alpha}\lambda_{\beta}\left\Vert g^{(\gamma)}\right\Vert \lambda_{\mu}\lambda_{\nu}\right]$
and $\delta(u,\gamma_{1})\delta(d,\gamma_{2})$. However, in the second
approach $A_{4}$ is associated with submatrix element
\begin{equation}
\left[\lambda_{\alpha}\lambda_{\beta}u\left\Vert g^{(\gamma)}(1,2)\right\Vert \lambda_{\mu}\lambda_{\nu}d\right]\\
={\displaystyle \sum_{\gamma_{1}\gamma_{2}}}\left[j_{\alpha},j_{\beta},\gamma,d\right]^{1/2}\left\{ \begin{array}{ccc}
j_{\mu} & j_{\nu} & d\\
\gamma_{1} & \gamma_{2} & \gamma\\
j_{\alpha} & j_{\beta} & u\end{array}\right\}\left[\lambda_{\alpha}\lambda_{\beta}\left\Vert g^{(\gamma_{1}\gamma_{2})}\right\Vert \lambda_{\mu}\lambda_{\nu}\right]R_{\alpha\beta\mu\nu}(1,2). \label{antigzusb}
\end{equation}

Here $R_{\alpha\beta\mu\nu}(1,2)$ is a radial integral of a radial function $g(r_{1},r_{2})$ in the basis of $|nlj)$ functions. In (\ref{antigzusb}) the coupled two-electron wave functions \begin{equation}
\left\vert n_{i}\lambda_{i}n_{j}\lambda_{j}um_{u}\right\rangle =\displaystyle \sum_{{\scriptstyle m_{\lambda_{i}}m_{\lambda_{j}}}}\left[\begin{array}{ccc}
\lambda_{i} & \lambda_{j} & u\\
m_{\lambda_{i}} & m_{\lambda_{j}} & m_{u}\end{array}\right]\left\vert n_{i}\lambda_{i}m_{i}n_{j}\lambda_{j}m_{\lambda_{j}}\right\rangle \label{coupl}\end{equation}
are used to determine the matrix element of $g_{m_{\gamma}}^{(\gamma)}$.
Note that in the first approach (\ref{ebscem}) uncoupled wave functions $\left\vert n_{\alpha}\lambda_{\alpha}m_{\alpha}n_{\beta}\lambda_{\beta}m_{\beta}\right\rangle \equiv\left\vert n_{\alpha}\lambda_{\alpha}m_{\alpha}\right\rangle \left\vert n_{\beta}\lambda_{\beta}m_{\beta}\right\rangle $
were employed. Bellow the indices $b$ and $z$ denote the quantities
which have been obtained in the first (\ref{ebscem}) and the second (\ref{ezschem}) approaches, respectively.

The coupling scheme of the block $T$ is developed to consider the
order of creation and annihilation operators ($A_{2}$). Let us study
this problem in a more detail. We collect the terms of operator $G$
taking into account on how many subshells of equivalent electrons
creation and annihilation operators act on. Then we can write\begin{equation}
G={\displaystyle \sum_{i}}G_{i}+{\displaystyle \sum_{i<j}}G_{ij}+{\displaystyle \sum_{i<j<k}}G_{ijk}+{\displaystyle \sum_{i<j<k<l}}G_{ijkl}.\label{opersum1}\end{equation}
Operators $G_{i},$ $G_{ij},$ $G_{ijk},$ $G_{ijkl}$ act in the
space of the states of one, two, three and four subshells, respectively.
Indices $i,j,k,l$ numerate the subshells in $\left\vert \Psi^{N}\right\rangle $
the operator $G$ acts on. In our study, the sums in (\ref{opersum1})
run in a way that $i<j<k<l$. The placing (arrangement) of creation
and annihilation operators in (\ref{opersum1}) follows the suggestions
of \cite{mg20}: first of all, operators $a_{m_{\lambda}}^{(\lambda)}$
and $\widetilde{a}_{m_{\lambda}}^{(\lambda)}$ which act on the same
subshell are collected side by side; secondly, operators $a_{m_{\lambda}}^{(\lambda)}$
and $\widetilde{a}_{m_{\lambda}}^{(\lambda)}$ acting on the first
(second, third) subshell of many-electron wave function are situated
to the left of the ones acting on the second (third, fourth) subshell.
Each operator in (\ref{opersum1}) is given in the coupled form \begin{equation}
G_{i...l}=\sum_{J_{1}J_{2}\,\varrho_{e}}\ ^{x}G_{s\varrho_{e}\qquad m_{\gamma}}^{(J_{1}J_{2})(\gamma)}\,^{x}g(s,\varrho_{e},J_{1},J_{2},\gamma).\label{gform1}\end{equation}
$^{x}G_{s\varrho_{e}\qquad m_{\gamma}}^{(J_{1}J_{2})(\gamma)}$ denotes
the irreducible tensorial product of the operators $a_{m_{\lambda}}^{(\lambda)}$
and $\widetilde{a}_{m_{\lambda}}^{(\lambda)}$ with the intermediate
ranks $J_{1}$, $J_{2}$ and with the resulting rank $\gamma$. $T$
(see diagram $A_{3}$) defines the coupling scheme of the irreducible
tensorial product. The factor $^{x}g(s,\varrho_{e},J_{1},J_{2},\gamma)$
(the diagrams $A_{3}$ and $A_{4}$) includes the submatrix elements
of $g_{i}^{\left(\gamma_{i}\right)}$ and the recoupling coefficients
arising while making the tensorial product. In (\ref{gform1}) superscript
$x$ prescribes in which approach the quantities are obtained. Argument $x=b,z$
indicates the first and second approaches, respectively; $s$ and
$\varrho_{e}$ characterize the set $\{n_{\alpha}\lambda_{\alpha}n_{\beta}\lambda_{\beta}n_{\mu}\lambda_{\mu}n_{\nu}\lambda_{\nu}\}$
of quantum numbers in (\ref{gform1}); $s$ indicates the number of
subshells the operator $G$ acts on. We collect together the terms
of operator $G$ with definite $s$ into the groups. Operators $G_{s\varrho\qquad m_{\gamma}}^{(J_{1}J_{2})(\gamma)}$
of a particular group connect exclusively the configuration states
$\left\langle \ldots N_{1},\ldots N_{2},\ldots N_{3},...N_{4}\ldots\right\vert $
and $\left\vert \ldots N_{1}^{\prime},\ldots N_{2}^{\prime},\ldots N_{3}^{\prime},...N_{4}^{\prime}\ldots\right\rangle $
with the specific electron occupation numbers $N_{i}$ and $N_{i}^{\prime}$,
\textit{i.e.}, $\{\delta_{1}=N_{1}^{\prime}-N_{1},$ $\delta_{2}=N_{2}^{\prime}-N_{2},$
$\delta_{3}=N_{3}^{\prime}-N_{3},\,\delta_{4}=N_{4}^{\prime}-N_{4}\}$.
Furthermore, the terms of each group $\varrho$ are collected into
the subgroups ($e$ numerates different subgroups). Each subgroup
$\varrho_{e}$ is characterized by the following sets of the quantum
numbers: $st1=\{n_{\alpha}\lambda_{\alpha}n_{\beta}\lambda_{\beta}n_{\nu}\lambda_{\nu}n_{\mu}\lambda_{\mu}\},$
$st2=\{n_{\beta}\lambda_{\beta}n_{\alpha}\lambda_{\alpha}n_{\mu}\lambda_{\mu}n_{\nu}\lambda_{\nu}\},$
$st3=\{n_{\alpha}\lambda_{\alpha}n_{\beta}\lambda_{\beta}n_{\mu}\lambda_{\mu}n_{\nu}\lambda_{\nu}\},$
$st4=\{n_{\beta}\lambda_{\beta}n_{\alpha}\lambda_{\alpha}n_{\nu}\lambda_{\nu}n_{\mu}\lambda_{\mu}\}.$
The subgroups differ the sets with distinct collections of $n_{\alpha}\lambda_{\alpha},$
$n_{\beta}\lambda_{\beta},n_{\nu}\lambda_{\nu},n_{\mu}\lambda_{\mu}$
quantum numbers. Note that the terms with $st1$ ($st2$) and $st3$
($st4$) describe the direct and exchange interactions, correspondingly.
Creation and annihilation operators with fixed $\varrho_{e}$ compose
the irreducible tensorial product $G_{s\varrho_{e}\qquad m_{\gamma}}^{(J_{1}J_{2})(\gamma)}$
(the diagram $A_{2}$). In general, the factor $^{x}g(s,\varrho_{e},J_{1},J_{2},\gamma)$
has four terms associated with $st1,st2,st3,st4$. However, due to
the symmetry properties of atomic interactions, the term in $^{x}g(s,\varrho_{e},J_{1},J_{2},\gamma)$
corresponding to the set $st1$ ($st3$) is equal to the term described
by $st2(st4)$.

For convenience, the expressions of $^{x}G_{s\varrho_{e}\qquad m_{\gamma}}^{(J_{1}J_{2})(\gamma)}$
and $^{x}g(s,\varrho_{e},J_{1},J_{2},\gamma)$ for $s=2,3,4$ are
collected in Tables \ref{tb:Table1}-\ref{tb:Table3}. The expressions for the operator $G$ when
it acts on one subshell ($n_{\alpha}\lambda_{\alpha}=n_{\beta}\lambda_{\beta}=n_{\mu}\lambda_{\mu}=n_{\nu}\lambda_{\nu},$
$s=1$) can be found everywhere (see, for example, in \cite{RK84},
\cite{grcff}, \cite{mg20}), therefore they are not presented in
this paper.

Let us concentrate now on the study of the tensorial part $^{x}G_{s\varrho_{e}\qquad m_{\gamma}}^{(J_{1}J_{2})(\gamma)}$
of $G$ for $s>1$. Decreasing a number of expressions which should
be written for $G_{s\varrho_{e}\qquad m_{\gamma}}^{(J_{1}J_{2})(\gamma)}$,
we explore the convention $T_{\alpha\beta}(b_{1},b_{2},b_{3},b_{4})$
to describe the tensorial products of creation and annihilation operators.
In $T_{\alpha\beta}$ the $i$-th operator is equal to $a^{\left(\lambda\right)}$
or $\widetilde{a}^{\left(\lambda\right)}$ if the argument $b_{i}=-1$
or $1$. For instance, $T_{21}(b_{1},b_{2},b_{3},b_{4})$ describes
three operators $G_{2\varrho_{e}\qquad m_{\gamma}}^{(J_{1}J_{2})(\gamma)}$
when $\varrho=1,2$ and $5$ (see Table \ref{tb:Table1}). It is important to note
that in our study in the case of three and four subshells the operators
$^{x}G_{2\varrho_{e}\qquad m_{\gamma}}^{(J_{1}J_{2})(\gamma)}$ are
identical for $x=b$ and $x=z$. Only for two-subshell case, the
operators $^{b}G_{2\varrho_{e}\qquad m_{\gamma}}^{(J_{1}J_{2})(\gamma)}$
and $^{z}G_{2\varrho_{e}\qquad m_{\gamma}}^{(J_{1}J_{2})(\gamma)}$
differ. More exactly, this takes place only if $\{\delta_{1},\delta_{2}\}=$
$\{-1,1\}$, $\{1,-1\}$, (\textit{i.e.}, when $\varrho=3,4$) and
then $e=1,2$ for fixed $\varrho$. Thus, for the remaining cases
index $e$ is redundant and it is dropped out in $^{x}G_{s\varrho_{e}\qquad m_{\gamma}}^{(J_{1}J_{2})(\gamma)}$
and $^{x}g(s,\varrho_{e},J_{1},J_{2},\gamma)$.


\begin{table}
\caption{The quantities for generation of the expressions for the operator $\mathnormal{G}$ in two-subshell case.}
\label{tb:Table1}
\begin{tabular}{|c|c|c|c|c|c|}
\hline 
$(\delta_{1},\delta_{2})$ & $\varrho$ & $e$ & $\:^{x}G_{2\varrho_{e}\:\:\:\:\:\:\: m_{\gamma}}^{(J_{1}J_{2})(\gamma)}$ & $\:^{b}g_{2\varrho_{e}}$ & $\:^{z}g_{2\varrho_{e}}$\tabularnewline
\hline
\hline 
$(0,0)$ & $1$ & $1$ & $T_{21}(-1,1,-1,1)$ & $\begin{array}{l}
(-1)^{j_{1}+j_{2}+J_{2}}\\
\times((-1)^{\gamma_{2}}b_{1212}D+\phi_{1})\\
+(b_{1221}D_{2}^{b}+\phi_{2})\end{array}$ & $-Z_{1212}D_{21}^{z}$\tabularnewline
\hline
\hline 
$(-2,2)$ & $2$ & $1$ & $T_{21}(-1,-1,1,1)$ & $-b_{1122}D$ & $Z_{1122}D_{22}^{z}$\tabularnewline
\hline
\hline 
$(2,-2)$ & $5$ & $1$ & $T_{21}(1,1,-1,-1)$ & $\overline{^{b}g}_{22_{1}}$ & $\overline{^{z}g}_{22_{1}}$\tabularnewline
\hline
\hline 
$(-1,1)$ & $3$ & $1$ & $\:^{x}T_{22}(-1,-1,1,1)$ & $\begin{array}{l}
(-1)^{j_{2}+J_{2}+\gamma_{1}+\gamma_{2}}(b_{1112}\\
\times\delta_{J_{1}\gamma_{2}}P+\phi_{1})\end{array}$ & $Z_{1112}\delta_{J_{1}u}D_{23}^{z}$\tabularnewline
\hline
\cline{3-3} \cline{4-4} \cline{5-5} \cline{6-6} 
\multicolumn{1}{|c|}{} & \multicolumn{1}{c|}{} & $2$ & $\:^{x}T_{23}(-1,-1,1,1)$ & $\begin{array}{l}
(-1)^{j_{1}+J_{2}+\gamma}(b_{1222}\\
\times\delta_{J_{1}\gamma_{2}}P|_{j_{1}\leftrightarrow j_{2}}+\phi_{1})\end{array}$ & $\begin{array}{l}
Z_{1222}\delta_{J_{1}d}\\
\times D_{23}^{z}|_{j_{1}\leftrightarrow j_{2},u\leftrightarrow d}\end{array}$\tabularnewline
\hline
\hline 
$(1,-1)$ & $4$ & $1$ & $\:^{x}T_{22}(-1,1,1,-1)|_{b\leftrightarrow z}$ & $(-1)^{j_{1}-J_{2}}\:\overline{^{b}g}_{23_{1}}$ & $(-1)^{j_{1}-J_{2}}\:\overline{^{z}g}_{23_{1}}$\tabularnewline
\cline{3-3} \cline{4-4} \cline{5-5} \cline{6-6} 
\multicolumn{1}{|c|}{} & \multicolumn{1}{c|}{} & $2$ & $\:^{x}T_{23}(1,-1,-1,1)|_{b\leftrightarrow z}$ & $(-1)^{j_{2}-J_{2}}\:\overline{^{b}g}_{23_{2}}$ & $(-1)^{j_{2}-J_{2}}\:\overline{^{z}g}_{23_{2}}$\tabularnewline
\hline
\end{tabular}
\end{table}

Operators $G_{2\varrho_{e}\qquad m_{\gamma}}^{(J_{1}J_{2})(\gamma)}$
(Table \ref{tb:Table1}) which act on two subshells ($s=2$) are described by tensorial
products with the following coupling schemes: \begin{equation}
T_{21}(-1,1,-1,1):=\left[T^{(J_{1})}\left(\lambda_{1},\widetilde{\lambda}_{1}\right)\times T^{(J_{2})}\left(\lambda_{2},\widetilde{\lambda}_{2}\right)\right]_{m_{\gamma}}^{(\gamma)},\label{duvien}\end{equation}
 \begin{equation}
^{b}T_{22}(-1,-1,1,1):=\left[\left[a^{\left(\lambda_{1}\right)}\times T^{(J_{1})}\left(\lambda_{1},\widetilde{\lambda}_{1}\right)\right]^{\left(J_{2}\right)}\times\widetilde{a}^{\left(\lambda_{2}\right)}\right]_{m_{\gamma}}^{(\gamma)},\label{dudub}\end{equation}
 \begin{equation}
^{b}T_{23}(-1,-1,1,1):=\left[a^{\left(\lambda_{1}\right)}\times\left[T^{(J_{1})}\left(\lambda_{2},\widetilde{\lambda}_{2}\right)\times\widetilde{a}^{\left(\lambda_{2}\right)}\right]^{\left(J_{2}\right)}\right]_{m_{\gamma}}^{(\gamma)},\label{dutrysb}\end{equation}
 \begin{equation}
^{z}T_{22}(-1,-1,1,1):=\left[\left[T^{(J_{1})}\left(\lambda_{1},\lambda_{1}\right)\times\widetilde{a}^{\left(\lambda_{1}\right)}\right]^{\left(J_{2}\right)}\times\widetilde{a}^{\left(\lambda_{2}\right)}\right]_{m_{\gamma}}^{(\gamma)},\label{duduz}\end{equation}
 \begin{equation}
^{z}T_{23}(-1,-1,1,1):=\left[a^{\left(\lambda_{1}\right)}\times\left[a^{\left(\lambda_{2}\right)}\times T^{(J_{1})}\left(\widetilde{\lambda}_{2},\widetilde{\lambda}_{2}\right)\right]^{\left(J_{2}\right)}\right]_{m_{\gamma}}^{(\gamma)}.\label{dutrysz}\end{equation}


\begin{table}
\caption{The quantities for generation of the expressions for the operator $\mathnormal{G}$ in three-subshell case.}
\label{tb:Table2}
\begin{tabular}{|c|c|c|c|c|}
\hline 
$(\delta_{1},\delta_{2},\delta_{3})$ & $\varrho$ & $\:^{x}G_{3\varrho\:\:\:\:\:\:\: m_{\gamma}}^{(J_{1}J_{2})(\gamma)}$ & $\:^{b}g_{3\varrho}$ & $\:^{z}g_{3\varrho}$\tabularnewline
\hline
\hline 
$(0,-1,1)$ & $2$ & $T_{31}(-1,1,-1,1)$ & $\begin{array}{l}
-(b_{1213}EN+\phi_{1})\\
+(b_{1231}\delta_{J_{1}\gamma_{1}}TN+\phi_{2})\end{array}$ & $\begin{array}{l}
-Z_{1213}\\
\times D_{31}^{z}|_{u\leftrightarrow d}\end{array}$\tabularnewline
\hline
\hline 
$(0,1,-1)$ & $1$ & $T_{31}(1,-1,1,-1)$ & $(-1)^{J_{1}}\:\overline{^{b}g}_{32}$ & $(-1)^{J_{1}}\:\overline{^{z}g}_{32}$\tabularnewline
\hline
\hline 
$(-2,1,1)$ & $4$ & $T_{31}(-1,-1,1,1)$ & $(b_{1123}D_{3}^{b}+\phi_{1})$ & $Z_{1123}D_{33}^{z}$\tabularnewline
\hline
\hline 
$(2,-1,-1)$ & $3$ & $T_{31}(1,1,-1,-1)$ & $\overline{^{b}g}_{34}$ & $\overline{^{z}g}_{34}$\tabularnewline
\hline
\hline 
$(-1,1,0)$ & $5$ & $T_{32}(-1,1,-1,1)$ & $\begin{array}{c}
(-1)^{j_{2}-j_{3}}\\
\times((-1)^{\gamma_{2}}b_{1323}PN+\phi_{1})\\
+(b_{1332}D_{2}^{b}+\phi_{2})\end{array}$ & $Z_{1323}D_{32}^{z}$\tabularnewline
\hline
\hline 
$(1,-1,0)$ & $6$ & $T_{32}(1,-1,-1,1)$ & $(-1)^{J_{2}}\:\overline{^{b}g}_{35}$ & $(-1)^{J_{2}}\:\overline{^{z}g}_{35}$\tabularnewline
\hline
\hline 
$(-1,-1,2)$ & $7$ & $T_{32}(-1,-1,1,1)$ & $(b_{1233}PN+\phi_{1})$ & $Z_{1233}D_{22}^{z}$\tabularnewline
\hline
\hline 
$(1,1,-2)$ & $8$ & $T_{32}(1,1,-1,-1)$ & $\overline{^{b}g}_{37}$ & $\overline{^{z}g}_{37}$\tabularnewline
\hline
\hline 
$(1,0,-1)$ & $9$ & $T_{33}(1,-1,1,-1)$ & $\begin{array}{l}
(-1)^{j_{1}+J_{1}-J_{2}}\\
\times\left.^{b}g_{31}\right|_{j_{1}\leftrightarrow j_{2}}\end{array}$ & $\begin{array}{l}
(-1)^{j_{1}+J_{1}-J_{2}}\\
\times\left.^{z}g_{31}\right|_{j_{1}\leftrightarrow j_{2}}\end{array}$\tabularnewline
\hline
\hline 
$(-1,0,1)$ & $10$ & $T_{33}(-1,1,-1,1)$ & $\begin{array}{l}
(-1)^{j_{1}+J_{1}-J_{2}}\\
\times\left.^{b}g_{32}\right|_{j_{1}\leftrightarrow j_{2}}\end{array}$ & $\begin{array}{l}
(-1)^{j_{1}+J_{1}-J_{2}}\\
\times\left.^{z}g_{32}\right|_{j_{1}\leftrightarrow j_{2}}\end{array}$\tabularnewline
\hline
\hline 
$(-1,2,-1)$ & $11$ & $T_{33}(-1,1,1,-1)$ & $\begin{array}{l}
(-1)^{j_{1}+J_{1}-J_{2}}\\
\times\left.^{b}g_{33}\right|_{j_{1}\leftrightarrow j_{2}}\end{array}$ & $\begin{array}{l}
(-1)^{j_{1}+J_{1}-J_{2}}\\
\times\left.^{z}g_{33}\right|_{j_{1}\leftrightarrow j_{2}}\end{array}$\tabularnewline
\hline
\hline 
$(1,-2,1)$ & $12$ & $T_{33}(1,-1,-1,1)$ & $\begin{array}{l}
(-1)^{j_{1}+J_{1}-J_{2}}\\
\times\left.^{b}g_{34}\right|_{j_{1}\leftrightarrow j_{2}}\end{array}$ & $\begin{array}{l}
(-1)^{j_{1}+J_{1}-J_{2}}\\
\times\left.^{z}g_{34}\right|_{j_{1}\leftrightarrow j_{2}}\end{array}$\tabularnewline
\hline
\end{tabular}
\end{table}

Here we have used the definition\begin{equation}
T^{(J)}\left(\lambda_{i},\widetilde{\lambda}_{j}\right):=\left[a^{\left(\lambda_{i}\right)}\times\widetilde{a}^{\left(\lambda_{j}\right)}\right]^{\left(J\right)}.\label{ttlambd}\end{equation}

For three subshells ($s=3$), operators $G_{3\varrho\qquad m_{\gamma}}^{(J_{1}J_{2})(\gamma)}$
(Table \ref{tb:Table2}) are given by the following types of tensorial products:\begin{equation}
T_{31}(-1,1,1,-1):=\left[\left[T^{(J_{1})}\left(\lambda_{1},\widetilde{\lambda}_{1}\right)\times\widetilde{a}^{\left(\lambda_{2}\right)}\right]^{\left(J_{2}\right)}\times a^{\left(\lambda_{3}\right)}\right]_{m_{\gamma}}^{(\gamma)},\label{trysvien}\end{equation}
 \begin{equation}
T_{32}(-1,1,-1,1):=\left[T^{(J_{1})}\left(\lambda_{1},\widetilde{\lambda}_{2}\right)\times T^{(J_{2})}\left(\lambda_{3},\widetilde{\lambda}_{3}\right)\right]_{m_{\gamma}}^{(\gamma)},\label{trystrys}\end{equation}
and \begin{equation}
T_{33}(1,-1,1,-1):=\left[\left[\widetilde{a}^{\left(\lambda_{1}\right)}\times T^{(J_{1})}\left(\lambda_{2},\widetilde{\lambda}_{2}\right)\right]^{\left(J_{2}\right)}\times a^{\left(\lambda_{3}\right)}\right]_{m_{\gamma}}^{(\gamma)}.\label{trysdu}\end{equation}

Finally, the four-subshell case ($s=4$) is presented by the tensorial
product \begin{equation}
T_{41}(-1,-1,1,1):=\left[\left[T^{(J_{1})}\left(\lambda_{1},\lambda_{2}\right)\times\widetilde{a}^{\left(\lambda_{3}\right)}\right]^{\left(J_{2}\right)}\times\widetilde{a}^{\left(\lambda_{4}\right)}\right]_{m_{\gamma}}^{(\gamma)}.\label{keturivien}\end{equation}
and operators $G_{4\varrho\qquad m_{\gamma}}^{(J_{1}J_{2})(\gamma)}$
in Table \ref{tb:Table3}. Note again, that in (\ref{duvien})-(\ref{keturivien})
operators $T_{\alpha\beta}(b_{1},b_{2},b_{3},b_{4})$ describe all
types (in the sense of coupling scheme) of the tensorial products
used in the present paper. The expressions for operators $^{x}G_{s\varrho_{e}\qquad m_{\gamma}}^{(J_{1}J_{2})(\gamma)}$
are easily obtained from (\ref{duvien})-(\ref{keturivien}) and Tables
\ref{tb:Table1}-\ref{tb:Table3}. When a consecutive coupling of the resulting moments of subshells
in many-electron wave function ${\left\vert n_{a}\lambda_{a}^{N_{a}}\Lambda_{a}\, n_{b}\lambda_{b}^{N_{b}}\Lambda_{b}...n_{k}\lambda_{k}^{N_{k}}\Lambda_{k}\,\left(\Lambda_{ab}...\right)\;\Lambda M_{\Lambda}\right\rangle }$
is used, the formulas for matrix elements of $^{x}G_{s\varrho_{e}\qquad m_{\gamma}}^{(J_{1}J_{2})(\gamma)}$
can be immediately found from general expressions given in \cite{mg20}.


\begin{table}
\caption{The quantities for generation of the expressions for the operator $\mathnormal{G}$ in four-subshell case.}
\label{tb:Table3}
\begin{tabular}{|c|c|c|c|c|}
\hline 
$(\delta_{1},\delta_{2},\delta_{3},\delta_{4})$ & $\varrho$ & $\:^{x}G_{4\varrho\:\:\:\:\:\:\: m_{\gamma}}^{(J_{1}J_{2})(\gamma)}$ & $\:^{b}g_{4\varrho}$ & $\:^{z}g_{4\varrho}$\tabularnewline
\hline
\hline 
$(-1,-1,1,1)$ & $1$ & $T_{41}(-1,-1,1,1)$ & $\begin{array}{l}
-(b_{1243}DV+\phi_{1})\\
+(-1)^{j_{2}+j_{3}}\\
\times((-1)^{\gamma_{1}}b_{2143}D_{41}^{b}+\phi_{2})\end{array}$ & $-Z_{1243}D_{41}^{z}$\tabularnewline
\hline
\hline 
$(1,1,-1,-1)$ & $4$ & $T_{41}(1,1,-1,-1)$ & $\overline{^{b}g}_{41}$ & $\overline{^{z}g}_{41}$\tabularnewline
\hline
\hline 
$(-1,1,-1,1)$ & $2$ & $T_{41}(-1,1,-1,1)$ & $\begin{array}{l}
-(b_{3142}D_{41}^{b}+\phi_{1})\\
+(b_{3124}\delta_{J_{1}\gamma_{2}}D_{42}^{b}\\
+\phi_{2})\end{array}$ & $\begin{array}{l}
(-1)^{j_{1}+j_{2}+J_{1}+1}\\
\times Z_{3142}\\
\times D_{42}^{z}|_{j_{1}\leftrightarrow j_{2}}\end{array}$\tabularnewline
\hline
\hline 
$(1,-1,1,-1)$ & $6$ & $T_{41}(1,-1,1,-1)$ & $\overline{^{b}g}_{42}$ & $\overline{^{z}g}_{42}$\tabularnewline
\hline
\hline 
$(1,-1,-1,1)$ & $3$ & $T_{41}(1,-1,-1,1)$ & $\begin{array}{l}
(-1)^{j_{1}+j_{3}}\\
\times((-1)^{\gamma_{1}}b_{3241}DV+\phi_{1})\\
+(-1)^{j_{1}-j_{2}}\\
\times((-1)^{\gamma_{2}}b_{3214}\delta_{J_{1}\gamma_{2}}D_{42}^{b}\\
+\phi_{2})\end{array}$ & $Z_{3241}D_{42}^{z}$\tabularnewline
\hline
\hline 
$(-1,1,1,-1)$ & $5$ & $T_{41}(-1,1,1,-1)$ & $\overline{^{b}g}_{43}$ & $\overline{^{z}g}_{43}$\tabularnewline
\hline
\end{tabular}
\end{table}

Consider now in detail the factor $^{x}g(s,\varrho_{e},J_{1},J_{2},\gamma)$
(\ref{gform1}). When $x=b$ (the first approach), we obtain \[
^{b}g(s,\varrho_{e},J_{1},J_{2},\gamma)=\displaystyle \sum_{\gamma_{1}\gamma_{2}}((^{b}D_{s\varrho_{e}}(\gamma,\lambda_{\alpha},\lambda_{\beta},\lambda_{\nu},\lambda_{\mu},\gamma_{1},\gamma_{2},J_{1},J_{2})\]
\[\times\, b(\gamma,\lambda_{\alpha},\lambda_{\beta},\lambda_{\nu},\lambda_{\mu},\gamma_{1},\gamma_{2})+\phi_{1})+(^{b}E_{s\varrho_{e}}(\gamma,\lambda_{\alpha},\lambda_{\beta},\lambda_{\nu},\lambda_{\mu},\gamma_{1},\gamma_{2},J_{1},J_{2})\]
\begin{equation}
\times b(\gamma,\lambda_{\alpha},\lambda_{\beta},\lambda_{\mu},\lambda_{\nu},\gamma_{1},\gamma_{2})+\phi_{2})).\label{ggbb}\end{equation}
The expressions for the special cases of $^{b}g(s,\varrho_{e},J_{1},J_{2},\gamma)$
are given in the fifth column of Table \ref{tb:Table1} and the fourth column of
Tables \ref{tb:Table2}, \ref{tb:Table3}. In the case of (\ref{gcr})
\begin{equation}
b_{\alpha\beta\nu\mu}\equiv b(\gamma,\lambda_{\alpha},\lambda_{\beta},\lambda_{\nu},\lambda_{\mu},\gamma_{1},\gamma_{2}):=\frac{1}{2}\left[\frac{j_{\alpha},j_{\beta}}{\gamma_{1},\gamma_{2}}\right]^{1/2}
\left[\lambda_{\alpha}\lambda_{\beta}\left\Vert g^{(\gamma_{1}\gamma_{2})}\right\Vert \lambda_{\mu}\lambda_{\nu}\right] R_{\alpha\beta\mu\nu}(1,2).\label{b1234}\end{equation}

The factors $^{b}D_{s\varrho_{e}}$
and $^{b}E_{s\varrho_{e}}$ arrive due to the recoupling procedures
described previously. The explicit expressions of these factors are
obtained by using the relations in Table \ref{tb:Table4} dealing with the following
ones: \begin{equation}
K(x,y,\gamma_{x},\gamma_{y},\gamma_{xy},z):=\left[\gamma_{x},z\right]^{1/2}\left\{ \begin{array}{lll}
z & x & \gamma_{y}\\
\gamma_{x} & \gamma_{xy} & y\end{array}\right\} \label{kaks}\end{equation}
and\begin{equation}
L(x_{1},x_{2},y_{1},y_{2},\gamma_{x},\gamma_{y},\gamma_{xy},z_{1},z_{2}):=\left[\gamma_{x},\gamma_{y},z_{1},z_{2}\right]^{1/2}\left\{ \begin{array}{lll}
x_{1} & x_{2} & \gamma_{x}\\
y_{1} & y_{2} & \gamma_{y}\\
z_{1} & z_{2} & \gamma_{xy}\end{array}\right\} .\label{lax}\end{equation}


\begin{table}
\caption{The relations for determination of the recoupling coefficients.}
\label{tb:Table4}
\begin{tabular}{|l|l|}   \hline   Two-subshell & Formula \\ \hline $D_{2}^{b}$ & $\delta (J _{1},\gamma _{1})\,\delta (J _{2},\gamma _{2})\Delta (\gamma _{1},\gamma _{2},\gamma )$ \\ \hline $D_{21}^{z}$ & $L(j_{1},j_{2},j_{1},j_{2},u,d,\gamma,J_{1},J_{2})$ \\ \hline $D_{22}^{z}$ & $\delta (J _{1},u)\,\delta (J _{2},d)\Delta (u,d,\gamma )$ \\ \hline $D_{23}^{z}$ & $(-1)^{\gamma +j _{1}+j _{2}+J _{1}}K(j_{1},j_{2},d,u,\gamma,J_{2})$ \\ \hline $D$ & $L(j_{1},j_{2},j_{1},j_{2},\gamma_{1},\gamma_{2},\gamma,J_{1},J_{2})$ \\ \hline $P$ & $K(j_{1},j_{2},\gamma_{1},\gamma_{2},\gamma,J_{2})$ \\ \hline\hline   Three-subshell &  \\ \hline $D_{3}^{b}$ & $\left( -1\right) ^{j _{1}+j _{3}+\gamma _{1}+\gamma _{2}}K(j_{1},j_{3},\gamma_{1},\gamma_{2},\gamma,J_{2})$ \\ & $\times K(j_{1},J_{2},\gamma_{2},j_{1},j_{2},J_{1})$ \\ \hline $D_{31}^{z}$ & $(-1)^{J _{1}+\gamma +j _{2}-j _{3}}K(j _{1},j _{3},u,d,\gamma ,J _{2})$ \\  & $\times K(j_{1},J_{2},d,j_{1},j_{2},J_{1})$ \\ \hline $D_{32}^{z}$ & $L\left( j _{1},j_{3},j _{2},j _{3},u,d,\gamma ,J _{1},J _{2}\right) $ \\ \hline $D_{33}^{z}$ & $(-1)^{\gamma+u+j_{2}+j_{3}}K(j _{2},j _{3},d,u,\gamma ,J _{2})\delta (J _{1},u)$ \\ \hline $TN$ & $(-1)^{\gamma+\gamma_{1}+j_{2}+j_{3}} K(j_{2},j_{3},\gamma_{2},\gamma_{1},\gamma,J_{2})$ \\ \hline $EN$ & $(-1)^{-\gamma_{1}+j_{1}-j_{3}}K(j_{1},j_{3},\gamma_{1},\gamma_{2},\gamma,J_{2})$ \\ & $\times K(j_{1},J_{2},\gamma_{2},j_{1},j_{2},J_{1})$ \\ \hline $PN$ & $(-1)^{1-J_{2}}L(j_{1},j_{3},j_{2},j_{3},\gamma_{1},\gamma_{2},\gamma,J_{1},J_{2})$ \\ \hline\hline Four-subshell &  \\ \hline $D_{41}^{b}$ & $(-1)^{j _{1}-j _{4}-\gamma _{1}+\gamma}K(j _{1},j _{4},\gamma_{2},\gamma_{1},\gamma,J_{2})$ \\  & $\times K(j_{1},J_{2},\gamma_{1},j_{2},j_{3},J_{1})$ \\ \hline $D_{42}^{b}$ & $(-1)^{j _{3}+j _{4}-\gamma _{1}}K(j _{3},j _{4},\gamma _{1},\gamma _{2},\gamma ,J _{2})$ \\ \hline $D_{41}^{z}$ & $(-1)^{d+\gamma +J _{1}}\,\delta (J _{1},u)\,K(j _{3},j _{4},d,u,\gamma ,J _{2})$ \\ \hline $D_{42}^{z}$ & $(-1)^{d+u+\gamma+1}K(j_{1},j_{4},d,u,\gamma,J_{2})$ \\  & $\times K(j_{1},J_{2},u,j_{2},j_{3},J_{1})$ \\ \hline $DV$ & $(-1)^{-\gamma+j_{3}-j_{4}+J_{1}}K(j_{2},j_{4},\gamma_{2},\gamma_{1},\gamma,J_{2})$ \\ & $\times K(j_{2},J_{2},\gamma_{1},j_{1},j_{3},J_{1})$ \\ \hline
\end{tabular}
\end{table}

To give more compact expression for $^{b}g(s,\varrho_{e},J_{1},J_{2},\gamma),$
we have introduced the notations $\phi_{1}$ and $\phi_{2}$ for the
second terms in the brackets in (\ref{ggbb}). The formula for the term
$\phi_{1}$ ($\phi_{2}$) linked to $st2$ ($st4$) is obtained from
the expression of the first term in the brackets associated with $st1$
($st3$) by replacing $b_{\alpha\beta\nu\mu}$ ($b_{\alpha\beta\mu\nu}$)
with $b_{\beta\alpha\mu\nu}$ ($b_{\beta\alpha\nu\mu}$), interchanging
$\gamma_{1}$ and $\gamma_{2}$ ($\gamma_{1}\leftrightarrow\gamma_{2}$),
and multiplying the obtained formula by the factor $(-1)^{\gamma_{1}+\gamma_{2}-\gamma}\equiv\omega$.
Furthermore, in Tables the notation $\overline{^{b}g}(s,\varrho_{e}^{\prime},J_{1},J_{2},\gamma)$
implies that the expression for $^{b}g(s,\varrho_{e},J_{1},J_{2},\gamma)$
is found from one of $^{b}g(s,\varrho_{e}^{\prime},J_{1},J_{2},\gamma)$
when $b_{\alpha\beta\nu\mu}$ is replaced with $b_{\mu\nu\beta\alpha}$
and the obtained formula is multiplied by $(-1)^{j_{\alpha}+j_{\beta}+j_{\nu}+j_{\mu}-\gamma_{1}-\gamma_{2}}$.
Note that for the atomic interactions when $g_{ij}=g_{ji}$, the term
$\phi_{1}$ ($\phi_{2}$) is equal to the term associated with $st1$
($st3$). When some effective interaction $^{eff}g_{_{12}\,\, m_{\gamma}}^{(\gamma)}$
is studied, the expression for $G$ can be found by replacing $b(\gamma,\lambda_{\alpha},\lambda_{\beta},\lambda_{\nu},\lambda_{\mu},\gamma_{1},\gamma_{2})$
with the factor $\left[\frac{j_{\alpha},j_{\beta}}{\gamma_{1},\gamma_{2}}\right]^{1/2}\left[\lambda_{\alpha}\lambda_{\beta}\left\Vert ^{eff}g^{(\gamma)}\right\Vert \lambda_{\mu}\lambda_{\nu}\right]$.
Finally, for obtaining the expression of $^{b}g(s,\varrho_{e},J_{1},J_{2},\gamma)$
in the case of the antisymmetric matrix element (\ref{eq:antap}),
the factor $b(\gamma,\lambda_{\alpha},\lambda_{\beta},\lambda_{\nu},\lambda_{\mu},\gamma_{1},\gamma_{2})$
of the term with $st1$ in (\ref{ggbb}) must be replaced by\[
b_{\alpha\beta\nu\mu}^{A}:=\frac{1}{2}(b(\gamma,\lambda_{\alpha},\lambda_{\beta},\lambda_{\nu},\lambda_{\mu},\gamma_{1},\gamma_{2})-\displaystyle \sum_{p_{1}, p_{2}}(-1)^{j_{\alpha}-j_{\beta}-\gamma_{2}-p_{2}}\left[j_{\alpha},j_{\beta},p_{1},p_{2}\right]^{1/2}\]
 \begin{equation}
\times\left\{ \begin{array}{ccc}
j_{\nu} & j_{\alpha} & \gamma_{1}\\
j_{\beta} & j_{\mu} & \gamma_{2}\\
p_{1} & p_{2} & \gamma\end{array}\right\} b(\gamma,\lambda_{\alpha},\lambda_{\beta},\lambda_{\mu},\lambda_{\nu},p_{1},p_{2})).\label{aggbb}\end{equation}

In addition, the terms corresponding to $st2,$ $st3$ and $st4$
must be dropped in (\ref{ggbb}).

In the second approach, the general formula for $^{x}g$ $(x=z)$
is presented as follows:
\begin{equation}
^{z}g(s,\varrho_{e},J_{1},J_{2},\gamma):=\,\displaystyle \sum_{u\, d}\,^{z}E_{s\varrho_{e}}(J_{1},J_{2},\gamma,\lambda_{\alpha},\lambda_{\beta},\lambda_{\nu},\lambda_{\mu},u,d)
Z(\gamma,\lambda_{\alpha},\lambda_{\beta},\lambda_{\nu},\lambda_{\mu},u,d).\label{ggzz}
\end{equation}
The expressions of the factors $\,^{z}E_{s\varrho_{e}}(J_{1},J_{2},\gamma,\lambda_{\alpha},\lambda_{\beta},\lambda_{\nu},\lambda_{\mu},u,d)\ $
which represent the recoupling coefficients are given in the last
column of Tables \ref{tb:Table1}-\ref{tb:Table3}. The second factor on the right side of (\ref{ggzz})
is expressed as
\begin{equation}
Z_{\alpha\beta\nu\mu}\equiv Z(\gamma,\lambda_{\alpha},\lambda_{\beta},\lambda_{\nu},\lambda_{\mu},u,d):=z_{\alpha\beta\nu\mu}+\eta(\alpha,\beta,u)z_{\beta\alpha\nu\mu}+\eta(\mu,\nu,d)z_{\alpha\beta\mu\nu}+\eta(\alpha,\beta,u)\eta(\mu,\nu,d)z_{\beta\alpha\mu\nu},\label{cczz}
\end{equation}
where $\eta(x,y,f):=(1-\delta(x,y))(-1)^{j_{x}+j_{y}-f+1}$. In the
case of (\ref{gcr}),
\begin{equation}
z_{\alpha\beta\nu\mu}\equiv z(\gamma,\lambda_{\alpha},\lambda_{\beta},\lambda_{\nu},\lambda_{\mu},u,d):=\frac{1}{2}(-1)^{j_{\mu}+j_{\nu}-d}\left[u/\gamma\right]^{1/2}
\left[\lambda_{\alpha}\lambda_{\beta}u\left\Vert g^{(\gamma)}(1,2)\right\Vert \lambda_{\mu}\lambda_{\nu}d\right].\label{zetmaz}
\end{equation}

Note that when $g_{ij}=g_{ji}$ the fourth (third) term in (\ref{cczz})
gives the same contribution as the first (second) one. For antisymmetric
coupled two-electron wave functions, the submatrix element of operator
$g_{\; m_{\gamma}}^{(\gamma)}$ is expressed as
\begin{equation}
\left[\lambda_{\alpha}\lambda_{\beta}u\left\Vert g^{(\gamma)}(1,2)\right\Vert \lambda_{\mu}\lambda_{\nu}d\right]_{A}=\left[1-(-1)^{j_{\mu}+j_{\nu}-\gamma}\left(\mu\leftrightarrow\nu\right)\right]\left[\lambda_{\alpha}\lambda_{\beta}u\left\Vert g^{(\gamma)}(1,2)\right\Vert \lambda_{\mu}\lambda_{\nu}d\right]\label{anampl}
\end{equation}
and
\begin{equation}
Z(\gamma,\lambda_{\alpha},\lambda_{\beta},\lambda_{\nu},\lambda_{\mu},u,d):=\frac{1}{4}\left(-1\right)^{j_{\mu}+j_{\nu}-d}\left[\frac{u}{\gamma}\right]^{1/2}\left[\lambda_{\alpha}\lambda_{\beta}u\left\Vert g^{(\gamma)}(1,2)\right\Vert \lambda_{\mu}\lambda_{\nu}d\right]_{A}.\label{czperantisim1}
\end{equation}
Furthemore, when studying the effective interaction 
\begin{equation}
z(\gamma,\lambda_{\alpha},\lambda_{\beta},\lambda_{\nu},\lambda_{\mu},u,d):=(-1)^{j_{\mu}+j_{\nu}-d}\left[u/\gamma\right]^{1/2}\left[\lambda_{\alpha}\lambda_{\beta}u\left\Vert ^{eff}g_{\; m_{\Gamma}}^{(\gamma)\Gamma}(1,2)\right\Vert \lambda_{\mu}\lambda_{\nu}d\right].\label{efesav}
\end{equation}
In Tables the notation $\overline{^{z}g}(\alpha,\varrho^{\prime},J_{1},J_{2},\gamma)$
means that the expression for $^{z}g(\alpha,\varrho,J_{1},J_{2},\gamma)$
is found from the expression for $\,^{z}g(\alpha,\varrho^{\prime},J_{1},J_{2},\gamma)|_{u\leftrightarrow d}$,
where $z_{\alpha\beta\nu\mu}$ is replaced with $z_{\mu\nu\beta\alpha}$
and the obtained formula is multiplied by the phase factor $(-1)^{j_{\alpha}+j_{\beta}+j_{\nu}+j_{\mu}-\gamma}.$ 
The explicite expressions of (\ref{b1234}), (\ref{efesav}) submatrix elements of Coulomb, magnetic and retardation interactions can be found, for instance, in \cite{ja73}, \cite{zr97}.


\section{Closed subshell cases}\label{sc:sc3}

Particularly simple expressions for the two-electron operator (\ref{ggr})
can be obtained when it acts on at least one closed subshell (say
$n_{\alpha}\lambda_{\alpha}^{2j_{\alpha}+1}$) of $\vert\Psi{}^{N}\rangle$.
Then, in the first approach from (\ref{ggr}) we obtain\[
^{b}G_{\alpha\beta\mu}=\displaystyle \sum_{{\scriptstyle n_{\beta}\lambda_{\beta}n_{\mu}\lambda_{\mu}n_{\alpha}\lambda_{\alpha}}}\hat{N}_{\alpha}T_{m_{\gamma}}^{(\gamma)}(\lambda_{\beta},\widetilde{\lambda}_{\mu})[j_{\alpha}]^{-1}{\displaystyle \sum_{\gamma_{1}\gamma_{2}}}\left[-[j_{\alpha}]^{1/2}\left(\delta_{\gamma_{1}0}\delta_{\gamma_{2}\gamma}b_{\alpha\beta\mu\alpha}+\delta_{\gamma_{1}\gamma}\delta_{\gamma_{2}0}b_{\beta\alpha\alpha\mu}\right)\right.\] 

\begin{equation}
+(-1)^{j_{\beta}+j_{\mu}+\gamma}[\gamma_{1},\gamma_{2}]^{1/2}\times\left.\left(\left\{ \begin{array}{ccc}
\gamma_{1} & \gamma_{2} & \gamma\\
j_{\mu} & j_{\beta} & j_{\alpha}\end{array}\right\} b_{\beta\alpha\mu\alpha}+\omega\left\{ \begin{array}{ccc}
\gamma_{1} & \gamma_{2} & \gamma\\
j_{\beta} & j_{\mu} & j_{\alpha}\end{array}\right\} b_{\alpha\beta\alpha\mu}\right)\right].\label{eq:3.1.5} 
\end{equation}

In the second approach
\begin{equation}
^{z}G_{\alpha\beta\mu}=2\displaystyle \sum_{{\scriptstyle n_{\beta}\lambda_{\beta}n_{\mu}\lambda_{\mu}n_{\alpha}\lambda_{\alpha}}}\hat{N}_{\alpha}T_{m_{\gamma}}^{(\gamma)}(\lambda_{\beta},\widetilde{\lambda}_{\mu})[j_{\alpha}]^{-1}{\displaystyle \sum_{ud}}(-1)^{\gamma+u+d+1}[u,d]^{1/2}\left\{ \begin{array}{ccc}
u & d & \gamma\\
j_{\mu} & j_{\beta} & j_{\alpha}\end{array}\right\} z_{\alpha\beta\mu\alpha}.\label{eq:3.1.6} 
\end{equation}

Here the operator $\hat{N}=-[\lambda]^{1/2}\left[a^{\left(\lambda\right)}\times\widetilde{a}^{\left(\lambda\right)}\right]^{\left(0\right)}$
has a submatrix element equal to $N$. In the case when $G$ acts
on two closed subshells (say $n_{\alpha}\lambda_{\alpha}^{2j_{\alpha}+1}$, $n_{\beta}\lambda_{\beta}^{2j_{\beta}+1}$),
we obtain
\begin{equation}
^{b}G_{\alpha\beta}=\displaystyle \sum_{{\scriptstyle n_{\alpha}\lambda_{\alpha}n_{\beta}\lambda_{\beta}}}\delta_{\gamma0}\hat{N}_{\alpha}\hat{N}_{\beta}[j_{\alpha},j_{\beta}]^{-1/2}{\displaystyle \sum_{\gamma_{1}\gamma_{2}}}\left\{ \delta_{\gamma_{1}0}\delta_{\gamma_{2}0}b_{\alpha\beta\beta\alpha}+(-1)^{j_{\alpha}+j_{\beta}}\right.\left.\delta_{\gamma_{1}\gamma_{2}}(-1)^{\gamma_{1}}[\gamma_{1}]^{1/2}b_{\alpha\beta\alpha\beta}\right\}.\label{eq:3.1.3} 
\end{equation}

\begin{equation}
^{z}G_{\alpha\beta}=-2\displaystyle \sum_{{\scriptstyle n_{\alpha}\lambda_{\alpha}n_{\beta}\lambda_{\beta}}}\delta_{\gamma0}\hat{N}_{\alpha}\hat{N}_{\beta}[j_{\alpha},j_{\beta}]^{-1}{\displaystyle \sum_{u}}\delta_{ud}[u]^{1/2}z_{\alpha\beta\alpha\beta}.\label{eq:3.1.4}\end{equation}
Finally, when $G$ acts on one closed subshell (say $n_{\alpha}\lambda_{\alpha}^{2j_{\alpha}+1}$),
we obtain\begin{equation}
^{b}G_{\alpha}=\displaystyle \sum_{{\scriptstyle n_{\alpha}\lambda_{\alpha}}}{\displaystyle \sum_{\gamma_{1}\gamma_{2}}}\hat{N}_{\alpha}(\hat{N}_{\alpha}-1)\delta_{\gamma0}\delta_{\gamma_{1}0}\delta_{\gamma_{2}0}[j_{\alpha}]^{-1}b_{\alpha\alpha\alpha\alpha}.\label{eq:3.1.1}\end{equation}

\begin{equation}
^{z}G_{\alpha}=\displaystyle \sum_{{\scriptstyle n_{\alpha}\lambda_{\alpha}}}\hat{N}_{\alpha}(\hat{N}_{\alpha}-1)\delta_{\gamma0}[j_{\alpha}]^{-2}{\displaystyle \sum_{u}}\delta_{ud}(-1)^{u}[u]^{1/2}z_{\alpha\alpha\alpha\alpha}.\label{eq:3.1.2}\end{equation}


\section{Conclusions}

A two-electron operator $G$ which describes the relativistic interactions
in atoms was considered in a coupled tensorial form in $jj$-coupling.
The second-quantization representation was used. A complete set of the expressions
when $G$ acts on two, three and four subshells (the largest number
of subshells the operator can act at the same time) of many-electron
wave function $\left\vert \Psi^{N}\right\rangle $ are presented in
a compact form. It allows easy generation of the formula for $G$
when the particular case is considered. Each expression is given
in such a structure that the calculation of the matrix elements of $G$ can
be performed into two separate tasks: the calculation of $N$-electron
spin-angular part presented by a submatrix element of the irreducible
tensorial product $G_{s\varrho\qquad m_{\gamma}}^{(J_{1}J_{2})(\gamma)}$
composed from creation and annihilation operators (the most computer
time-consuming part of calculations) and the determination of the
factors $^{x}g(s,\varrho_{e},J_{1},J_{2},\gamma)$. The factors $^{x}g(s,\varrho_{e},J_{1},J_{2},\gamma)$
do not depend on the resulting angular momentum of subshells $\Lambda_{a}$
of $\left\vert \Psi^{N}\right\rangle $, thus they can be determined
before the calculations of submatrix elements for $G_{s\varrho\qquad m_{\gamma}}^{(J_{1}J_{2})(\gamma)}$
and reduce the computation time of matrix elements.

In the present paper we apply the coupling schemes for $G_{s\varrho\qquad m_{\gamma}}^{(J_{1}J_{2})(\gamma)}$
which are very useful in the case of the consecutive order coupling
of the resulting momenta of subshells in $\left\vert \Psi^{N}\right\rangle $.
Then many-electron submatrix element of $G_{s\varrho\qquad m_{\gamma}}^{(J_{1}J_{2})(\gamma)}$
takes a very simple expression, \textit{i.e.}, it is expressed as sums
which run over only intermediate ranks ($J_{1},J_{2}$) of the products
of $6j$ and/or $9j$-symbols and submatrix elements of operators
acting in the space of states formed from a subshell of equivalent
electrons. In the present paper two forms of coupling schemes of angular
momenta of two-electron operator $G$ were studied. It enables us to
use uncoupled (the first approach), coupled (the second approach)
and antisymmetric two-electron wave functions in constructing coupled
tensorial form of the operator. The possibility to apply different
types of two-electron wave functions allows one to choose more optimal
ways of calculations. Note that the second approach is preferable for
the problems where several operators with different tensorial structures
are considered, for instance, in the formation of energy matrix for
the atomic Hamiltonian. In this case, Coulomb and Breit interactions
can be presented by a single operator $G$ with two-electron submatrix
element $\left[\lambda_{\alpha}\lambda_{\beta}u\left\Vert g^{(\gamma)}\right\Vert \lambda_{\mu}\lambda_{\nu}d\right]$
when $\gamma=0$, where $g^{(0)}=g_{Coulomb}^{(0)}+g_{Breit}^{(0)}$.
Then many-electron angular part can be determined as the submatrix
element of the unique operator $G_{s\varrho\qquad m_{\gamma}}^{(J_{1}J_{2})(\gamma)}$.
The first approach is more preferable when one seeks to
calculate the matrix elements of particular operator efficiently. In this approach,
the internal tensorial structure of the operator $g^{(\gamma)}$ is
directly involved (through the diagram $A_{4}$, the coupling scheme
$E_{b}$) into the expressions of many-electron matrix element of $G_{s\varrho\qquad m_{\gamma}}^{(J_{1}J_{2})(\gamma)}$
and recoupling coefficient (the diagram $A_{4}$).

It is important to note that the expressions of both approaches are
also applicable to the study of the operators representing some effective
interactions in atoms arising, for instance, in Atomic MBPT or Coupled
Cluster (CC) method.

The method to obtain the formulas for the operator $G$ developed
in \cite{frcffgaig}, \cite{grcff}, can be explained as
the combination of our first and second approaches. However, the methodology proposed
in our paper on coupling schemes to construct the irreducible tensorial
products of creation and annihilation operators allows us to find 
expressions for many-electron matrix element of the operator $G$ more simply
than in \cite{frcffgaig}.

\section*{Acknowledgments}
The study was partially funded by the Joint Taiwan-Baltic Research project.

\end{document}